\documentclass[11pt,a4paper]{article}
\usepackage{jheppub}
\usepackage{amsmath}
\usepackage{amssymb}
\usepackage{braket}
\usepackage{graphicx}
\usepackage{caption}
\usepackage{subcaption}

\newcommand{\mO}{\mathcal{O}}	
\newcommand{\mK}{\mathcal{M}}	
\newcommand{\mR}{\mathcal{R}}	
\newcommand{\mA}{\mathcal{A}}

\newcommand{\kb}{k}
\newcommand{\lb}{l}
\newcommand{\hb}{h}
\newcommand{\Xb}{X}
\newcommand{\be}{\begin{equation}}
\newcommand{\ee}{\end{equation}}

\title{Holographic Proof of the Quantum Null Energy Condition}

\author[a,b]{Jason Koeller}
\author[a,b]{and Stefan Leichenauer}

\affiliation[a]{Center for Theoretical Physics and Department of Physics,\\
University of California, Berkeley, CA 94720, U.S.A.} 
\affiliation[b]{Lawrence Berkeley National Laboratory, Berkeley, CA 94720, U.S.A.} 

\emailAdd{jkoeller@berkeley.edu}
\emailAdd{sleichen@berkeley.edu}

\abstract{We use holography to prove the Quantum Null Energy Condition (QNEC) at leading order in large-$N$ for CFTs and relevant deformations of CFTs in Minkowski space which have Einstein gravity duals. Given any codimension-2 surface $\Sigma$ which is locally stationary under a null deformation in the direction $k$ at the point $p$, the QNEC is a lower bound on the energy-momentum tensor at $p$ in terms of the second variation of the entropy to one side of $\Sigma$: $\langle T_{kk}\rangle \geq S''/2\pi \sqrt{h}$. In a CFT, conformal transformations of this inequality give results which apply when $\Sigma$ is not locally stationary. The QNEC was proven previously for free theories, and taken together with our result this provides strong evidence that the QNEC is a true statement about quantum field theory in general.
}
\begin{document}
\maketitle

\section{Introduction}\label{sec-introduction}

The Null Energy Condition (NEC), $T_{kk} \equiv T_{ij}k^ik^j \geq 0$,  is ubiquitous in classical physics as a signature of stable field theories. In General Relativity it underlies many results, such as the singularity theorems~\cite{Penrose:1964wq,HawEll,Wald} and area theorems~\cite{Hawking:1971tu,Bousso:2015mqa}. In AdS/CFT, imposing the NEC in the bulk has several consequences for the field theory at leading order in large-$N$, including the holographic c-theorems~\cite{Myers:2010tj,Myers:2010xs,Freedman:1999gp} and Strong Subadditivity of the covariant holographic entanglement entropy~\cite{Wall:2012uf}. Yet ultimately the NEC, interpreted as a local bound on the expectation value $\langle T_{kk} \rangle$, is known to fail in quantum field theory~\cite{Epstein:1965zza}.

The Quantum Null Energy Condition (QNEC) was proposed in~\cite{Bousso:2015mna} as a correction the NEC which holds true in quantum field theory. In the QNEC, $\langle T_{kk} \rangle$ at a point $p$ is bounded from below by a nonlocal quantity constructed from the von Neumann entropy of a region. Suppose we divide space into two regions, one of which we call $\mR$, with the dividing boundary $\Sigma$ passing through $p$. We compute the entropy of $\mR$, and consider the second variation of the entropy as $\Sigma$ is deformed in the null direction $k^i$ at $p$. Call this second variation $S''$ (a more careful construction of $S''$ is given in below in Section~\ref{sec-statement}). Then the QNEC states that
\be
\langle T_{kk} \rangle \geq \frac{\hbar}{2\pi \sqrt{h}} S'', 
\ee
where $\sqrt{h}$ is the determinant of the induced metric on $\Sigma$ at the location $p$.\footnote{In general, there may be ambiguities in the definition of $T_{kk}$ because of ``improvement terms." It is plausible that a similar ambiguity in the definition of $S$ leaves the QNEC unaffected by these issues~\cite{Casini:2014yca, Akers:2015bgh, Herzog:2014fra,Lee:2014zaa}.} The QNEC has its origins in quantum gravity: it arose as a consequence of the Quantum Focussing Conjecture (QFC), proposed in~\cite{Bousso:2015mna}, but is itself a statement about quantum field theory alone.

In~\cite{Bousso:2015wca}, the QNEC was proved for the special case of free (or superrenormalizable) bosonic field theories for certain surfaces $\Sigma$. Here we will prove the QNEC for a completely different class of field theories, namely those which have a good gravity dual, at leading order in the large-$N$ expansion. We will consider any theory obtained from such a large-$N$ UV CFT by a scalar relevant deformation. We will also assume that the bulk theory is an Einstein gravity theory, so that the leading order part of the entropy is given by the area of an extremal surface in the bulk in Planck units:
\begin{align}\label{HRT}
	S = \frac{A(m)}{4 G_{N}\hbar}\, ,
\end{align}
where $A(m)$ is the area of a bulk codimension-two surface $m$ which is homologous to $\mR$ and is an extremum of the area functional in the bulk~\cite{Ryu:2006ef,Ryu:2006bv,Hubeny:2007xt}. Computing that change in the extremal area as the surface $\Sigma$ is deformed is then a simple task in the calculus of variations.\footnote{There can be phase transitions in the holographic entanglement entropy where $S'$ is discontinuous at leading order in $N$. This happens when there are two extremal surfaces with areas that become equal at the phase transition. Since we are instructed to use the minimum of the two areas to compute the entropy, the entropy function is always concave in the vicinity of the phase transition. Therefore $S'' = -\infty$ formally, so the QNEC is satisfied. Thus it is sufficient to assume that no phase transitions are encountered in the remainder of the paper.} A key property is that the change in area of an extremal surface under deformations is due entirely to the near-boundary asymptotic region, where a general analytic computation is possible.

Our proof method involves tracking the motion of $m$ as $\Sigma$ is deformed. The ``entanglement wedge" proposal for the bulk region dual to $\mR$, together with bulk causality, suggests that $m$ should move in a spacelike way as we deform $\Sigma$ in our chosen null direction~\cite{Czech:2012bh,Headrick:2014cta}, and a theorem of Wall~\cite{Wall:2012uf} shows that this is, in fact, correct.\footnote{We would like to thank Zachary Fisher, Mudassir Moosa, and Raphael Bousso for discussions about the spacelike nature of these deformations, as well as bringing the theorem of~\cite{Wall:2012uf} to our attention.} We construct a bulk vector $s^\mu$ in the asymptotic bulk region which points in the direction of the deformation of $m$, and since $s^\mu$ is spacelike we have $s^\mu s_\mu \geq 0$. Holographically, $\langle T_{kk} \rangle$ is encoded in the near-boundary expansion of the bulk metric, and therefore enters into the expression for $s^\mu s_\mu$. We will see that the inequality $s^{\mu} s_{\mu} \geq 0$ is precisely the QNEC.\footnote{Relations between the boundary energy-momentum tensor and a coarse-grained entropy were studied using holography in \cite{Bunting:2015sfa}. The entropy we consider in this paper is the fine-grained von Neumann entropy.}

The remainder of the paper is organized as follows. In Section~\ref{sec-statement} we will give a careful account of the construction of $S''$ and the statement of the QNEC. In Section~\ref{sec-proof} we prove the QNEC at leading order in large-$N$ using holography. In Section~\ref{sec:Expn} we recall the asymptotic expansions of the bulk metric and extremal surface embedding functions that we will use for the rest of our proof. In Section~\ref{sec:spacelike} we discuss the fact that null deformations of $\Sigma$ on the boundary induce spacelike deformations of $m$ in the bulk and define the spacelike vector $s^\mu$. In Section~\ref{sec-der} we construct $s^\mu$ in the asymptotic region and calculate its norm, thereby proving the QNEC. Then in Section~\ref{sec-QNECconformal} we specialize to CFTs and examine the QNEC in different conformal frames. Finally, in Section~\ref{sec-future} we discuss the outlook on extensions of the proof and its ideas, as well as possible applications of the QNEC.


\paragraph{Notation}
Our conventions follow those described in footnote 5 of \cite{Hung:2011ta}. Letters from the second half of the Greek alphabet (\(\mu, \nu, \rho, \dots\)) label directions in the bulk geometry. Letters from the second half of the Latin alphabet (\(i,j,k,\dots\)) label directions in the boundary. Entangling surface directions in the boundary (or on a cutoff surface) are denoted by letters from the beginning of the Latin alphabet (\(a,b,c,\dots\)), while directions along the corresponding bulk extremal surface are labeled with the beginning of the Greek alphabet (\(\alpha,\beta,\gamma,\dots\)). We will often put an overbar on bulk quantities to distinguish them from their boundary counterparts, e.g., $\bar{h}(z=0) = h$. We neglect the expectation value brackets when we refer to the expectation value of the boundary stress tensor, i.e. \(T_{ij} \equiv \braket{T_{ij}}\). Boundary latin indices $i,j,k, \ldots$ are raised and lowered with the boundary metric \(\eta_{ij}\). Outside of the Introduction we set $\hbar=1$.


\section{Statement of the QNEC}\label{sec-statement}

\begin{figure}[t]
	\centering
	\begin{subfigure}{.49\textwidth}
	\includegraphics[width=\textwidth]{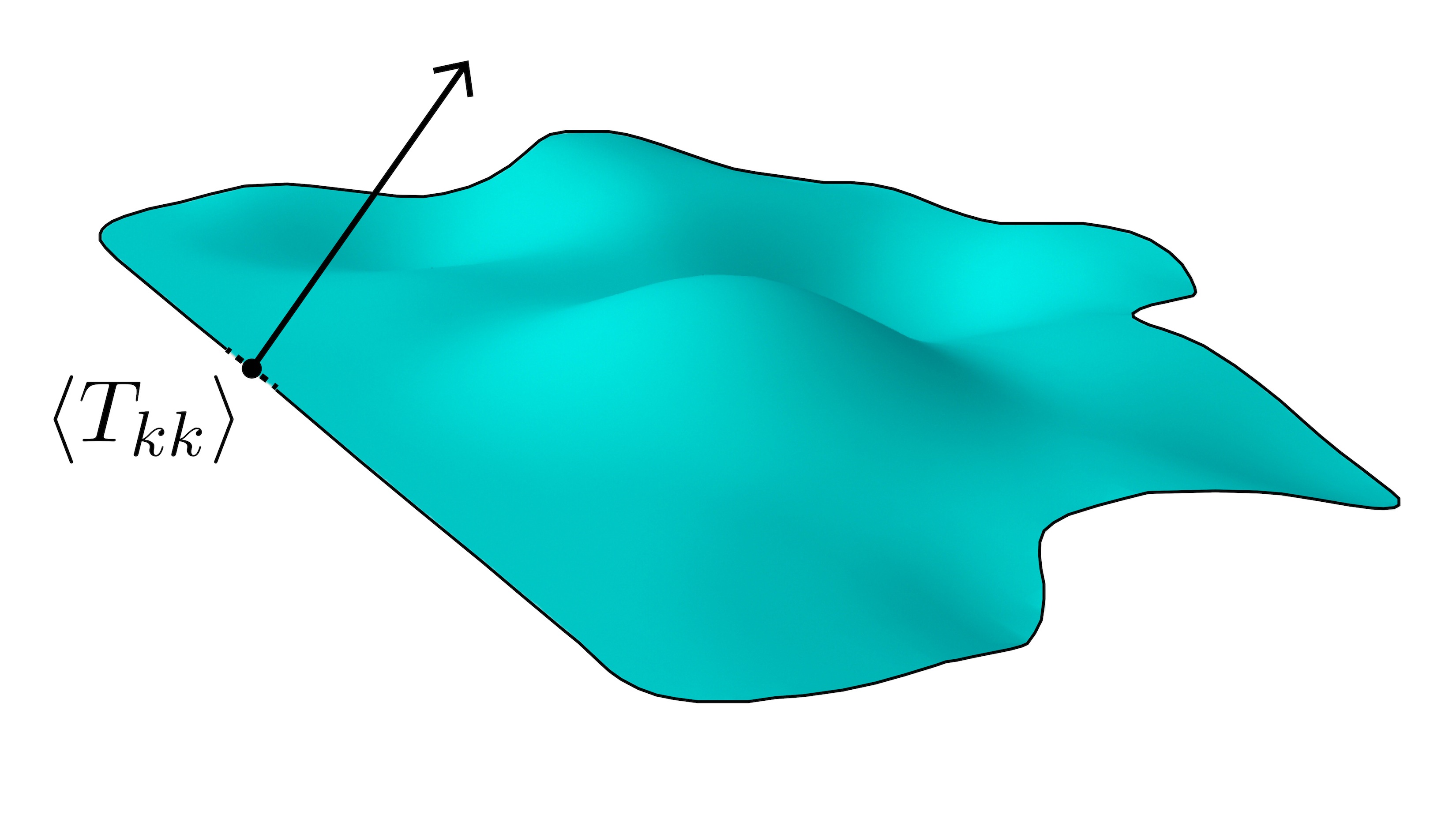}
	\end{subfigure}
	\begin{subfigure}{.49\textwidth}
	\includegraphics[width=\textwidth]{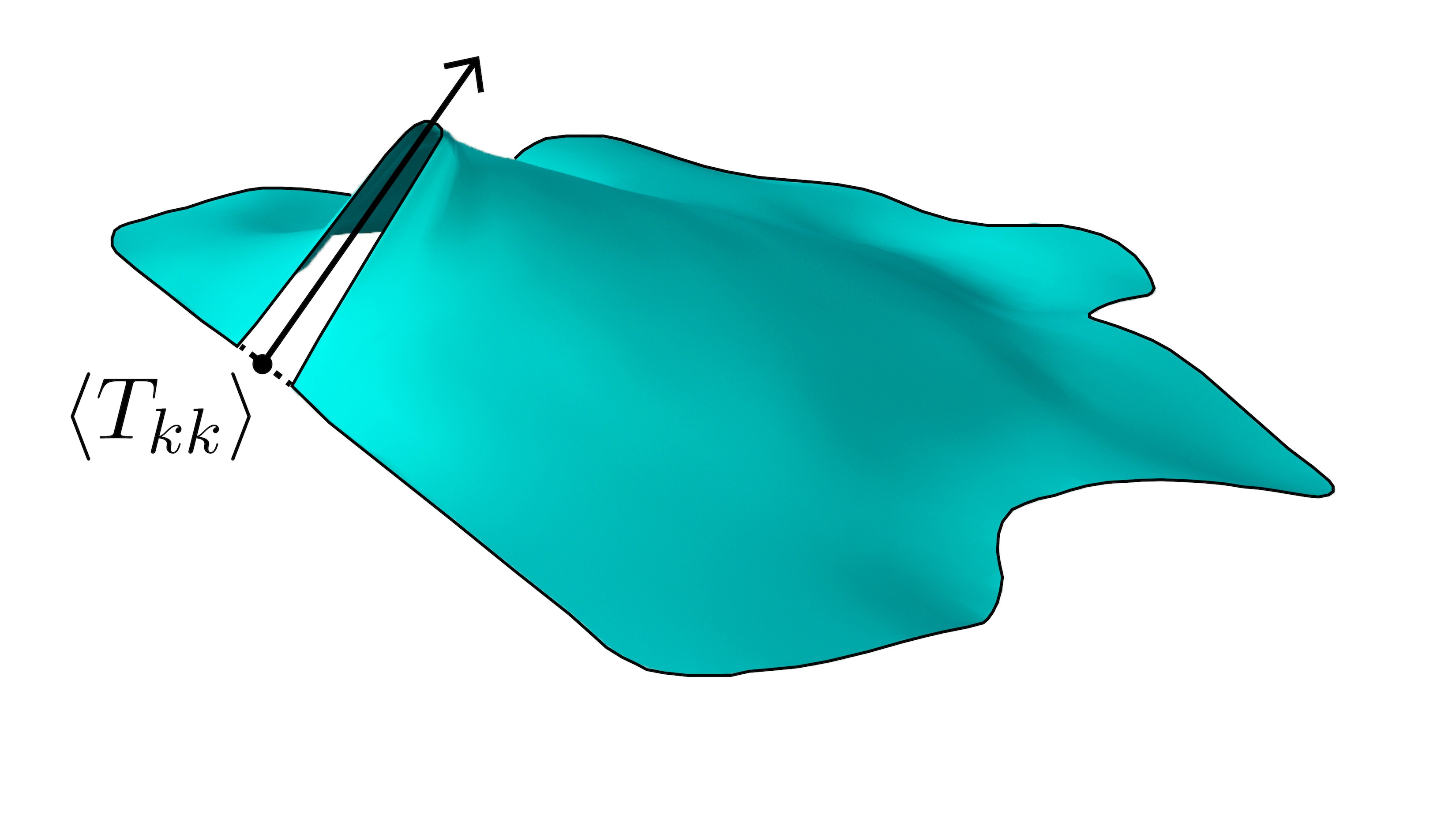}
	\end{subfigure}
	\caption{Here we show the region $\mR$ (shaded cyan) and the boundary $\Sigma$ (black border) before and after the null deformation. The arrow indicates the direction $\kb^i$, and $\braket{T_{kk}}$ is being evaluated at the location of the deformation. The dashed line indicates the support of the deformation.}\label{fig:deformation}
\end{figure}

In this section we will give a careful statement of the QNEC. Consider an arbitrary quantum field theory in $d$-dimensional Minkowski space. The QNEC is a pointwise lower bound on the expectation value of the null-null component of the energy-momentum tensor, $T_{kk} \equiv \langle T_{ij}\rangle \kb^i \kb^j$, in any given state. Let us choose a codimension-2 surface $\Sigma$ which contains the point of interest, is orthogonal to $\kb^i$, and divides a Cauchy surface into two regions. We can assign density matrices to the two regions of the Cauchy surface and compute their von Neumann entropies. In a pure state these two entropies will be identical, but we do not necessarily have to restrict ourselves to pure states. So choose one of the two regions, which we will call $\mR$ for future reference, and compute its entropy $S$. If we parameterize the surface $\Sigma$ by a set of embedding functions $\Xb^i(y)$ (where $y$ represents $d-2$ internal coordinates), then we can think of the entropy as a functional $S= S[\Xb^i(y)]$.

Our analysis is centered around how the functional $S[\Xb^i(y)]$ changes as the surface $\Sigma$ (and region $\mR$) is deformed.\footnote{Deformations of $\Sigma$ induce appropriate deformations of $\mR$~\cite{Bousso:2015mqa}.} Introducing a deformation $\delta \Xb^i(y)$, we can define variational derivatives of $S$ through the equation
\be\label{eq-variation}
\Delta S = \int dy \, \frac{\delta S}{\delta \Xb^i(y)} \delta \Xb^i(y) +\frac{1}{2}  \int dydy' \, \frac{\delta^2 S}{\delta \Xb^i (y) \delta \Xb^j(y')} \delta \Xb^i(y) \delta \Xb^j(y') + \cdots.
\ee
One might worry that the functional derivatives $\delta S /\delta \Xb^i(y)$, $\delta^2 S/\delta \Xb^i (y) \delta \Xb^j(y')$, and so on are unphysical by themselves because we cannot reasonably consider deformations of the surface on arbitrarily fine scales. But the functional derivatives are a useful tool for compactly writing the QNEC, and we can always integrate our expressions over some small region in order to get a physically well-defined statement. Below we will do precisely that to obtain the global version of the QNEC from the local version.

The QNEC relates $T_{kk}$ to the second functional derivative of the entropy under null deformations, i.e., the second term in \eqref{eq-variation} in the case where $\delta \Xb^i(y) = \kb^i(y)$ is an orthogonal null vector field on $\Sigma$. Let $\lambda$ be an affine parameter along the geodesics generated by $\kb^i(y)$; it will serve as our deformation parameter. Then we can isolate the second variation of the entropy by taking two derivatives with respect to $\lambda$:~\footnote{We use capital-\(D\) for ordinary derivatives to avoid any possible confusion with the \(S''\) notation. \(D\)-derivatives are defined by \eqref{eq-secondderivative}, while \(S''\) is defined by \eqref{eq-secondvariation}.}
\be\label{eq-secondderivative}
\frac{D^2 S}{D\lambda^2} = \int dydy' \, \frac{\delta^2 S}{\delta \Xb^i (y) \delta \Xb^j(y')} \kb^i(y) \kb^j(y').
\ee
It is important that $\kb^i(y)$ also satisfies a global monotonicity condition: the domain of dependence of $\mR$ must be either shrinking or growing under the deformation. In other words, the domain of dependence of the deformed region must either contain or be contained in the domain of dependence of the original region. By exchanging the role played by $\mR$ and its complement, we can always assume that the domain of dependence is shrinking. In this case the deformation has a nice interpretation in the Hilbert space in terms of a continuous tracing out of degrees of freedom. Then consider the following decomposition of the second variation of $S$ into a ``diagonal" part, proportional to a $\delta$-function, and an ``off-diagonal" part:
\be\label{eq-secondvariation}
 \frac{\delta^2 S}{\delta \Xb^i (y) \delta \Xb^j(y')} \kb^i(y) \kb^j(y') = S''(y) \delta(y-y') + (\text{off-diagonal}).
\ee
Our notation for the diagonal part, $S''(y)$, suppresses its dependence on the surface $\Sigma$, but it is still a complicated non-local functional of the $\Xb^i$. Because of the global monotonicity property of $\kb^i(y)$, one can show using Strong Subadditivity of the entropy that the ``off-diagonal" terms are non-positive~\cite{Bousso:2015mna}. We will make use of this property below to transition from the local to the global version of the QNEC.

For a generic point on a generic surface, $S''$ will contain cutoff-dependent divergent terms. It is easy to see why: the cutoff-dependent terms in the entropy are proportional to local geometric integrals on the entangling surface, and the second variation of such terms is present in $S''$.\footnote{Although it is the case that all of the cutoff-dependence in the second variation of the entropy is contained in the diagonal part, which we have called $S''$, it is still true that $S''$ contains finite terms as well. If it did not, the QNEC would be the same as the NEC.} By restricting the class of entangling surfaces we consider, we can guarantee that the cutoff-dependent parts of the entropy have vanishing second derivative. In the course of our proof (see section \ref{sec:Expn}), we will find that a sufficient condition to eliminate all cutoff-dependence in $S''$ is that $\kb_i K^i_{ab} =0$ in a neighborhood of the location where we wish to bound $T_{kk}$, where $K^i_{ab}$ is the extrinsic curvature tensor of $\Sigma$ (also known as the second fundamental form).\footnote{$K^i_{ab}$ is defined as $D_a D_b \Xb^i$, where $D_a$ is the induced covariant derivative on $\Sigma$.} The locality of this statement should be emphasized: away from the point where we wish to bound $T_{kk}$, $\Sigma$ can be arbitrary.

Finally, we can state the QNEC. When $\kb^i(y)$ satisfies the global monotonicity constraint and  $\kb_i K^i_{ab} = 0$ in a neighborhood of $y=y_0$, we have
\be\label{QNEC}
T_{kk} \geq \frac{1}{2\pi \sqrt{\hb}} S'' 
\ee
where $\sqrt{\hb}$ is the surface volume element of $\Sigma$ and all terms are evaluated at $y=y_0$. A few remarks are in order. In $d=2$, the requirement $\kb_i K^i_{ab} = 0$ is trivial. In that case we are also able to prove the stronger inequality
\be\label{QNEC'}
T_{kk} \geq \frac{1}{2\pi} \left[ S'' + \frac{6}{ c} \left(S'\right)^2\right].
\ee
Here $S' \equiv  \kb^i \delta S/\delta \Xb^i$ and $c$ is the central charge of the UV fixed point of the theory. This stronger inequality in $d=2$ is actually implied by the weaker one in the special case of a CFT by making use of the conformal transformation properties of the entropy~\cite{Wall:2011kb}, though here we will prove it even when the theory contains a relevant deformation. One can use similar logic in $d>2$ to generalize the statement of the QNEC when applied to a CFT. By Weyl transformation, we can transform a surface that has $\kb_i K^i_{ab} = 0$ to one where $\kb_i K^i_{ab}\hb^{ab} \neq 0$, though the trace-free part still vanishes. In that case, we will find
\be\label{QNECconformal}
T_{kk} - \mA^{(T)}_{kk} \geq \frac{1}{2\pi \sqrt{\hb}}\left[\left(S_{\rm fin}-\mA^{(S)}\right)'' + \frac{2\theta}{d-2} \left(S_{\rm fin} - \mA^{(S)}\right)'\right]
\ee
for CFTs in $d>2$, where $\theta \equiv -\kb_i K^i_{ab}\hb^{ab}$ is the expansion in the $\kb^i$ direction, and $\mA^{(T)}_{kk}$ and $\mA^{(S)}$ are anomalous shifts in $T_{kk}$ and $S$, respectively \cite{Graham:1999pm}. The two anomalies are both zero in odd dimensions, and $\mA^{(T)}_{\rm kk}$ is zero for global conformal transformations in Minkowski space. $\mA^{(S)}$ is a local geometric functional of $\Sigma$, and may be non-zero even when $\mA^{(T)}_{kk}$ vanishes. The finite part of the entropy appears in this equation because we are starting with the finite inequality \eqref{QNEC}. The Weyl-transformed surface violates the condition $\kb_i K^i_{ab} = 0$, so the divergent parts of the variation of $S$ do not automatically vanish. We will discuss this inequality in more detail in Section~\ref{sec-QNECconformal}.

Before continuing on with the proof of the QNEC, we should discuss briefly the integrated version. Suppose that $\kb_iK^i_{ab} = 0$ on all of $\Sigma$ (which we can always enforce by setting $\kb^{i}=0$ on some parts of $\Sigma$). Then we can integrate \eqref{QNEC} to obtain
\be\label{QNECglobal}
2\pi \int dy \sqrt{\hb}~ T_{kk}  \geq  \frac{D^2S}{D\lambda^2}.
\ee
Here we made use of \eqref{eq-secondderivative} and \eqref{eq-secondvariation}, and also the fact that the ``off-diagonal" terms in \eqref{eq-secondvariation} are non-positive \cite{Bousso:2015mna}. This is a global version of the QNEC, but it is actually equivalent to the local version. By considering the limiting case of a vector field $\kb^i(y)$ with support concentrated around $y=y_0$, we can obtain \eqref{QNEC} from \eqref{QNECglobal}.

\section{Proof of the QNEC}\label{sec-proof}

\subsection{Setup: Asymptotic Expansions}\label{sec:Expn}

Our proof of the QNEC relies on the form of the bulk metric and extremal surface near the AdS boundary. In this section, we review the Fefferman-Graham expansion of the bulk metric and the analogous expansion of the extremal embedding functions, recalling the relevant properties of each.

\paragraph{Metric Expansion}
We are only interested in QFTs formulated on $d$-dimensional Minkowski space. Through order \(z^{d}\), the asymptotic expansion of the metric near the AdS boundary takes the form
\begin{align}\label{metricExpn}
	ds^{2} = \frac{L^{2}}{z^{2}} \left( dz^{2} + \left[ f(z) \eta_{ij} + \frac{16 \pi G_{N}}{dL^{d-1}} z^{d} t_{ij} \right] dx^{i} dx^{j} + o(z^{d})\right) = G_{\mu\nu} dx^\mu dx^\nu.
\end{align}
Here $L$ is the AdS length, $f(z)$ only contains powers of $z$ less than $d$ (and possibly a term proportional to $z^d\log z$) and satisfies $f(0) =1$. The exact form of $f(z)$ will depend on the theory; in a CFT $f(z) =1$ but we are free to turn on relevant deformations which can modify it. We are assuming that only Poincare-invariant theories are being considered; this is why $\eta_{ij}$ is the only tensor appearing up to order $z^d$.

The tensor $t_{ij}$, defined by its appearance in \eqref{metricExpn} as the coefficient of $z^d$, is not necessarily the same as $T_{ij}$. In a CFT on Minkowski space they are equal, but in the presence of a relevant deformation one has to carefully define the renormalized energy-momentum tensor of the new theory.\footnote{See \cite{deHaro:2000xn} for example.} In particular, $t_{ij}$ may not vanish in the vacuum state of the deformed theory.  However, the difference $T_{ij} - t_{ij}$ is proportional to $\eta_{ij}$.\footnote{The difference should be proportional to the relevant coupling $\phi_0$, and dimensional analysis dictates that the only possibility is $\phi_0 \mO \eta_{ij}$ where $\mO$ is the relevant operator.} Therefore $t_{kk} = T_{kk}$, which is all we will need.

The \((d+1)\)-dimensional bulk metric is denoted by \(G_{\mu\nu}\), but we will also find it convenient to define the rescaled metric
 \begin{align}\label{smallg}
 	g_{\mu\nu} \equiv \frac{z^{2}}{L^{2}} G_{\mu\nu}.
 \end{align}

\paragraph{Embedding Functions}
The embedding of the (\(d-1\))-dimensional extremal surface \(m\) in the (\(d+1\))-dimensional bulk can be described by specifying the bulk coordinates as a function of $z$ and $(d-2)$ intrinsic coordinates $y^a$, \(\bar{X}^\mu = \bar{X}^\mu(y^{a},z)\). These functions are called the ``embedding functions."\footnote{Our index conventions are described at the end of the Introduction.} The induced metric on \(m\) is given by
\begin{align}\label{bigH}
	\bar{H}_{\alpha \beta} \equiv \partial_{\alpha } \bar{X}^\mu \partial_{\beta} \bar{X}^{\nu} G_{\mu\nu}[\bar{X}],
\end{align}
where \(G_{\mu\nu}\) is the bulk metric. Instead of \(\bar{H}_{\alpha\beta}\), it is often more convenient to use a rescaled surface metric:
\begin{align}\label{halphabetaDef}
	\bar{h}_{\alpha\beta} \equiv \partial_{\alpha} \bar{X}^\mu \partial_{\beta} \bar{X}^{\nu} g_{\mu\nu}[\bar{X}] = \frac{z^{2}}{L^{2}} \bar{H}_{\alpha\beta},
\end{align}
where \(g_{\mu\nu} = (z^{2}/L^{2})G_{\mu\nu}\) as defined above. Our internal coordinates for the surface are chosen so that \(\bar{H}_{az} = \bar{h}_{az} = 0\) and $\bar{X}^z = z$~\cite{Schwimmer:2008yh}.

The embedding functions satisfy an equation of motion coming from extremizing the total area. In terms of this induced metric, this can be written as \cite{Hung:2011ta}
\begin{align}\label{extremalSurfaceEquation}
	\frac{1}{\sqrt{\bar{H}}} \partial_{\alpha} \left(\sqrt{\bar{H}} \bar{H}^{\alpha \beta} \partial_{\beta} \bar{X}^\mu \right) + \bar{H}^{\alpha \beta} \Gamma^{\mu}_{\nu\sigma} \partial_{\alpha} \bar{X}^{\nu} \partial_{\beta} \bar{X}^{\sigma} = 0,
\end{align}
where \(\Gamma^{\mu}_{\nu\sigma}\) is the bulk Christoffel symbol constructed with the bulk metric \eqref{metricExpn} and \(\bar{H} \equiv \det{\bar{H}_{\alpha\beta}}\). The embedding functions have an asymptotic expansion near the boundary with a structure very similar to that of the bulk metric. There are two solutions, with the state-independent solution containing lower powers of \(z\) than the state-dependent solution. The state-independent solution only contains terms of lower order than $z^d$, and only depends on the state-independent part of the bulk metric \eqref{metricExpn}. If we only include the terms in \eqref{extremalSurfaceEquation} relevant for the terms of lower order than $z^d$, we find
\be
z^{d-1}\partial_z \left(z^{1-d}\sqrt{\bar{h}}\bar{h}^{zz} f \partial_z \bar{X}^i \right) + \partial_a \left(f \sqrt{\bar{h}}\bar{h}^{ab}\partial_b \bar{X}^i\right) = 0.
\ee
where \(\bar{h} \equiv \det{\bar{h}_{ab}}\).
The solution to this equation can be found algebraically order-by-order in $z$ up to $z^d$. The expansion reads
\be\label{embExpn}
	\bar{X}^i(y^{a},z) = \Xb^{i}(y^{a})+ \frac{1}{2(d-2)}z^2 K^i(y^a) + \cdots + \frac{1}{d}z^{d} \left(V^{i}(y^{a}) + W^i(y^a)\log z \right) + o(z^{d}).
\ee
Here $K^i$ is the trace of the extrinsic curvature tensor of the entangling surface $\Sigma$. Since the background geometry is flat, this can be written as
\be
K^i = \frac{1}{\sqrt{\hb}}\partial_a \left( \sqrt{\hb} \hb^{ab}\partial_b \Xb^i  \right).
\ee
The omitted terms ``$\cdots$" contain powers of $z$ between $2$ and $d$. In a CFT there would be only even powers, but with a relevant deformation odd or fractional powers are allowed depending the dimension of the relevant operator. These terms, as well as the logarithmic term $W^i$, are all state-independent,\footnote{They are only state-independent if there are no scalar operators of dimension $\Delta < d/2$. For the case of operators with $d/2 > \Delta > (d-2)/2$, see Appendix~\ref{stressTensorApp}.} and are local functions of geometric invariants of the entangling surface \cite{Hung:2011ta}. These geometric invariants are formed from contractions of the extrinsic curvature and its derivatives, and will vanish if the surface is flat: if $K^i$ vanishes in some neighborhood on the surface, then $\bar{X}^i = \Xb^i + V^i z^d/d$ satisfies the equation of motion up to that order in $z$. The logarithmic coefficient $W^i$ is only present in when $d$ is even for a CFT, but it may also show up in odd dimensions if relevant operators of particular dimensions are turned on. 

The state-dependent part of the solution starts at order \(z^{d}\), and the only term we have shown in \eqref{embExpn} is \(V^{i}\). We will find below that this term encodes the variation of the entropy that enters into the QNEC.

\paragraph{Extremal Surface Area Asymptotic Expansion}

With $\bar{H}_{\alpha\beta} = \partial_\alpha \bar{X}^\mu \partial_\beta \bar{X}^\nu G_{\mu\nu}$ the induced metric on the extremal surface, the area functional is
\be
A =  \int dzd^{d-2}y~ \sqrt{\bar{H}[\bar{X}]}.
\ee
We are interested in variations of the extremal area when the entangling surface $\Sigma$ is deformed. That is, when the boundary embedding functions $\Xb^i$ are varied. The variation of the area is not guaranteed to be finite: divergences will be regulated by a cutoff surface at $z=\epsilon$.  A straightforward exercise in the calculus of variations shows that
\be
\delta A = - \frac{L^{d-1}}{z^{d-1}}\int d^{d-2}y \left. \sqrt{\bar{h}}\frac{g_{ij} \partial_z \bar{X}^i }{\sqrt{1+g_{lm} \partial_z \bar{X}^l \partial_z \bar{X}^m}} \delta \bar{X}^j \right|_{z=\epsilon}.
\ee
Each factor in this expression (including $\delta \bar{X}^j$) should be expanded in powers of $z$ and evaluated at $z=\epsilon$. Making use of \eqref{metricExpn} and \eqref{embExpn}, we find
\be\label{eq-AExpn}
\frac{1}{L^{d-1}\sqrt{\hb}}\frac{\delta A}{\delta \Xb^i} = - \frac{1}{(d-2) \epsilon^{d-2}}  K_i + (\text{power law})  - W_i  \log \epsilon  -  V_i + (\text{finite state-independent}).
\ee
The most divergent term goes like $\epsilon^{2-d}$, and is the variation of the usual area-law term expected in any quantum field theory. The logarithmically divergent term is directly determined in terms of the logarithmic term in the expansion of the embedding functions in \eqref{embExpn}. The remaining terms, including both the lower-order power law divergences and the state-independent finite terms, are determined in terms of the ``$\cdots$"  of \eqref{embExpn}. Their precise form is not important, but our analysis later will depend on the fact that they are built out of local geometric data on \(\Sigma\), and that they vanish when $K^i_{ab}=0$ locally. That is, if $K^i_{ab}$ and its derivatives vanish at a point $y$, then these terms are zero at that point.

\paragraph{Elimination of Divergences}

Now we will illustrate that the condition $\kb_i K^i_{ab}  = 0$ in the neighborhood of a point is enough to remove divergences in $S''$.\footnote{In the remainder of proof we assume $\kb^i(y) \neq 0$. That is, we are only considering regions of the entangling surface which are actually being deformed.} First we note that the condition $\kb_{i} K_{ab}^i=0$ is robust under null deformations in the $\kb^i$ direction. That is, if it is satisfied initially then it remains satisfied for all values of $\lambda$. To see this, we use the identity\footnote{The extrinsic curvature is often defined as \(K^{i}_{ab} = \partial_{a}X^{l}\partial_{b}X^{m}\nabla_{l}h_{m}^{~i}\). ``Differentiating by parts'' and restricting to Minkowski space gives the first equality of equation \eqref{Kidentity}.}
\be\label{Kidentity}
\kb_i K_{ab}^i =\kb_i \partial_a\partial_b \Xb^i= - \partial_a \kb_i \partial_b \Xb^i
\ee
and take a $\lambda$-derivative to get
\be
\partial_\lambda(\kb_i K_{ab}^i) =- \partial_a \kb_i \partial_b \kb^i = - (\kb_i K^i_{ac}) \hb^{cd} (\kb_j K_{db}^j). 
\ee
For the last equality we used the fact that $\kb_i \partial_a \kb^i =0$, so the inner product could be evaluated by first projecting onto the tangent space of $\Sigma$. This shows that $\kb_i K_{ab}^i$ remains zero if it is initially zero, and so all of our remaining results hold even as we deform $\Sigma$.

We claim when $\kb_i K^i_{ab} = 0$ locally, the expansion \eqref{embExpn} reduces to 
\be\label{eq-Xexpansion}
\bar{X}^i(y,z) = \Xb^i(y) + B(y,z) \kb^i(y) + \frac{1}{d}V^i (y)z^d + o(z^d).
\ee 
Here $B(y,z)$ is a function which vanishes at $z=0$ and contains powers of $z$ less than $d$, and possibly a term proportional to $z^d \log z$. The nontrivial claim here is that the leading $z$ terms up to $z^d$ are all proportional to $\kb^i$. We will now prove this claim.

We know from the equations of motion that the terms of in the embedding function expansion at orders lower than $z^d$ are determined locally in terms of the geometry of the entangling surface. This means they can only depend on $\eta_{ij}$, $\partial_a \Xb^i$, $K_{ab}^i$, and finitely many derivatives of $K_{ab}^i$ in the directions tangent to $\Sigma$. If $K_{ab}^i$ is proportional to $\kb^i$, the same is true for its derivatives. To see this, we only need to show that $\partial_a \kb^i$ is proportional to $\kb^i$. Since $\kb^i$ is null, we have $\kb_i \partial_a \kb^i = 0$. Therefore $\partial_a\kb^i$ does not have any components in the null direction opposite to $\kb^i$ (which we will call $\lb^i$ below). We can also compute its components in the tangent directions:
\be
\partial_b \Xb^i \partial_a \kb_i = - \kb_i \partial_a\partial_b \Xb^i = -\kb_i K_{ab}^i = 0.
\ee
Hence $\partial_a \kb^i \propto \kb^i$, and so all of the tangent derivatives of $K^i_{ab}$ are proportional to $\kb^i$.

Now, one can check that if $K^i_{ab}$ and all of its derivatives are zero then \eqref{eq-Xexpansion} with $B=0$ solves the equation of motion up to order $z^d$. This means that at least one power of $K_{ab}^i$ (or its derivatives) must appear in each of the terms in the expansion of $\bar{X}^i$ of lower order than $z^d$ beyond zeroth order. But this means that at least one power of $\kb^i$ appears, and there are no tensors available to give nonzero contractions with $\kb^i$. Hence each of these terms must be proportional to $\kb^i$, and this is the claim of \eqref{eq-Xexpansion}. We emphasize that this expansion is valid in any state of the theory, even in the presence of a relevant deformation.

An analogous result holds for the expansion of the entropy variation, which means that \eqref{eq-AExpn} reduces to 
\be\label{eq-areavariation}
\frac{\delta A}{\delta \Xb^i(y)} = C(y,\epsilon) \kb_i(y) -L^{d-1} \sqrt{\hb(y)}\ V_i(y),
\ee
where \(C(y,\epsilon) \kb_{i}(y)\) represents the local terms (both divergent and finite) in \eqref{eq-AExpn}. But now we see that all divergent terms are absent in null variations of the area: by contracting \eqref{eq-areavariation} with $\kb^i$ we see that the only non-zero contribution is the finite state-dependent term $\kb^i V_i$.

\subsection{Proof Strategy: Extremal Surfaces are Not Causally Related}\label{sec:spacelike}

The QNEC involves the change in the von Neumann entropy of a region $\mR$ under the local transport of a portion of the entangling surface \(\Sigma\) along null geodesics (see Figure~\ref{fig:deformation}). The entropy $S(\mR)$ is computed as the area of the extremal surface $m(\mR)$ in the bulk, and so we need to analyze the behavior of extremal surfaces under boundary deformations. Our analysis is rooted in the following Fact: \emph{for any two boundary regions \(A\) and \(B\) with domain of dependence \(D(A)\) and \(D(B)\) such that \(D(A) \subset D(B)\), \(m(B)\) is spacelike- or null-separated from \(m(A)\).} This result is proved as theorem 17 in \cite{Wall:2012uf} and relies on the null curvature condition in the bulk, which in Einstein gravity is equivalent to the bulk (classical) NEC.\footnote{Strictly speaking, theorem 17 in \cite{Wall:2012uf} concludes that \(m(A)\) and \(m(B)\) are spacelike-separated, because the bulk null generic condition is assumed. However, special regions and special states will have null separation. For example, in the vacuum any region in \(d=2\) as well as spherical regions and half-spaces in arbitrary dimension have this property. This observation is used for spherical regions in section \ref{sec-QNECconformal}}

Even though this Fact can be proved based on properties of extremal area surfaces, it is useful to understand the intuition behind why it should be true. The idea, first advocated in~\cite{Czech:2012bh}, is that associated to the domain of dependence $D(A)$ of any region $A$ in the field theory should be a region $w(A)$ of the bulk, which in~\cite{Headrick:2014cta} was dubbed the ``entanglement wedge."  The extremal surface $m(A)$ is the boundary of the entanglement wedge. Consider two regions $A$ and $B$ satisfying $D(A) \subset D(B)$, and consider also the complement of region $B$, $\bar{B}$. Assume for simplicity that $m(B) = m(\bar{B})$. If some part of $m(A)$ were timelike-seperated from some part of $m(B)$, then that part of $m(A)$ would also be timelike-separated from $w(\bar{B})$. But the entanglement wedge proposal dictates that (unitary) field theory operators acting in $\bar{B}$ can influence the bulk state anywhere in $w(\bar{B})$, and so by bulk causality could influence the extremal surface $m(A)$ and thereby alter the entropy $S(A)$. But a unitary operator acting on $\bar{B}$ leaves the density matrix of $B$ invariant, and therefore also the density matrix of $A$, and therefore also $S(A)$. 

Based on this heuristic argument, one expects that a similar spacelike-separation property should exist for the boundaries of the entanglement wedges of $D(A)$ and $D(B)$ in any holographic theory, not just one where those boundaries are given by extremal area surfaces. For this reason, we are optimistic about the prospects for proving the QNEC using the present method beyond Einstein gravity, though we leave the details for future work.

\begin{figure}[t]
	\centering
	\includegraphics[width=.7\textwidth]{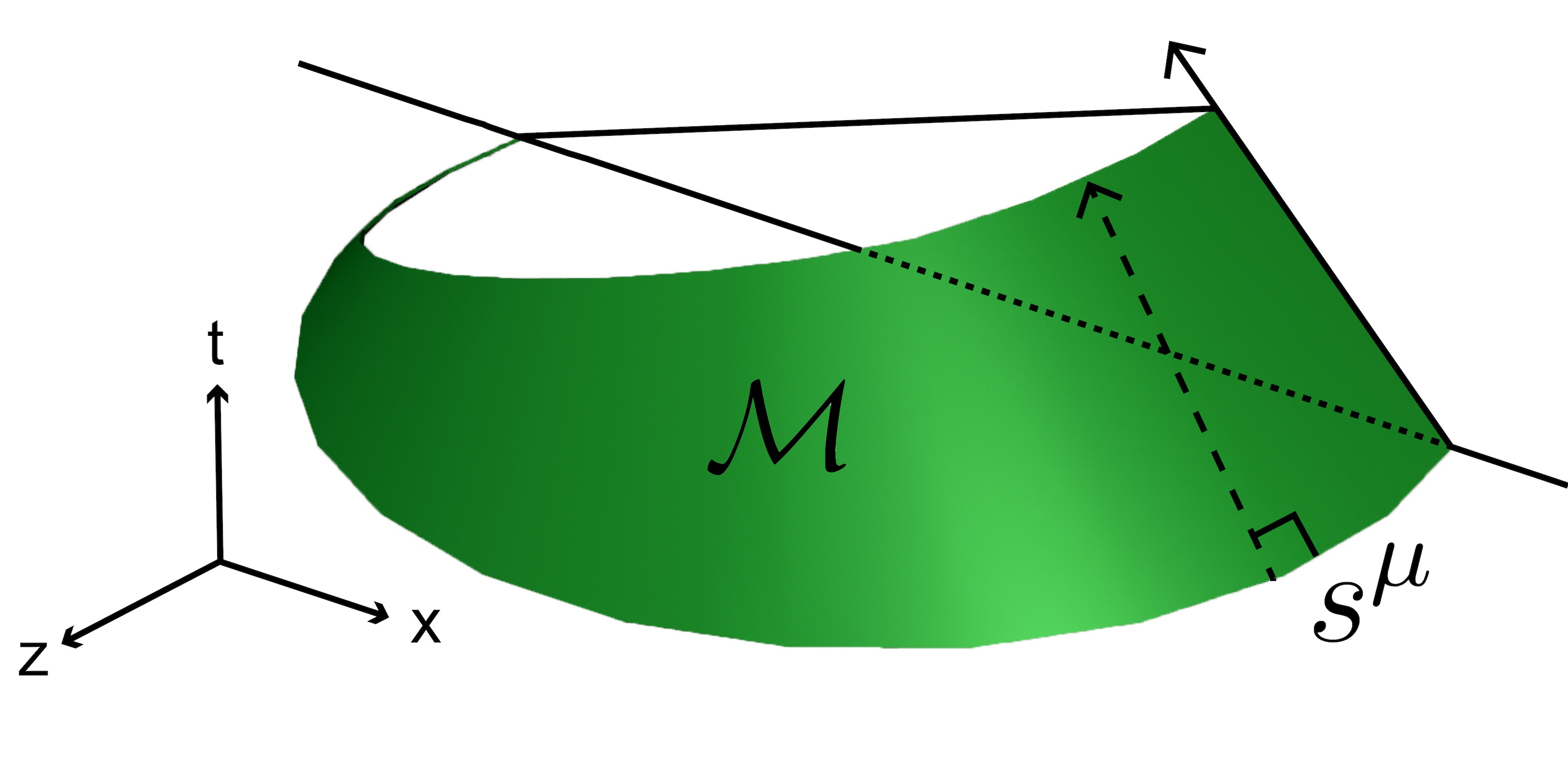}
	\caption{The surface $\mK$ in the bulk (shaded green) is the union of all of the extremal surfaces anchored to the boundary that are generated as we deform the entangling surface. The null vector $\kb^i$ (solid arrow) on the boundary determines the deformation, and the spacelike vector $s^\mu$ (dashed arrow) tangent to $\mK$ is the one we construct in our proof. The QNEC arises from the inequality $s^\mu s_\mu \geq 0$.}\label{fig:bulk}
\end{figure}

Let $\Sigma$ be the boundary of the region $\mR$. We consider deformations of $\Sigma$ by transporting it along orthogonal null geodesics generated by the orthogonal vector field $\kb^i$ on $\Sigma$, thus giving us a one-parameter family of entangling surfaces $\Sigma(\lambda)$ which bound the regions $\mR(\lambda)$, where $\lambda$ is an affine parameter of the deformation. We also obtain a one-parameter family of extremal surfaces $m(\mR(\lambda))$ in the bulk whose areas compute the entropies of the regions. Recall the global monotonicity constraint on $\kb^i$: we demand that the domain of dependence of $\mR(\lambda)$ is either shrinking or growing as a function of $\lambda$. In other words, we have either $D(\mR(\lambda_1)) \subset D(\mR(\lambda_2))$ or  $D(\mR(\lambda_2)) \subset D(\mR(\lambda_1))$ for every $\lambda_1 < \lambda_2$. Then, by the Fact quoted above, the union $\mK$ of all of the $m(\mR(\lambda))$ is an achronal hypersurface in the bulk (see Figure~\ref{fig:bulk}). That is, all tangent vectors on $\mK$ are either spacelike or null.\footnote{Part of theorem 17 in \cite{Wall:2012uf} is that the extremal surfaces associated to all the $\mR(\lambda)$ lie on a single bulk Cauchy surface. $\mK$ is just a portion of that Cauchy surface.} We will see that the QNEC is simply the non-negativity of the norm of a certain vector $s^\mu$ tangent to $\mK$: $g_{\mu\nu}s^\mu s^\nu \geq 0$.

Since $\mK$ is constructed as a one-parameter family of extremal surfaces (indexed by $\lambda$), we can take as a basis for its tangents space the vectors $\partial_a \bar{X}^\mu$, $\partial_z \bar{X}^\mu$, and $\partial_\lambda \bar{X}^\mu$. The first two are tangent to the extremal surface at each value of $\lambda$, while the third points in the direction of the deformation. One can check that the optimal inequality is given by choosing $s^\mu$ to be normal to the extremal surface $m(\mR)$. Thus we can simply define $s^\mu$ as the normal part of $\partial_\lambda \bar{X}^\mu$.

It turns out to be algebraically simplest to construct a null basis of vectors normal to the extremal surface at fixed $\lambda$ and then find the linear combination of them which is tangent to $\mK$. We begin with the null vectors $\kb^i$, $\lb^i$ on the boundary which are orthogonal to the entangling surface. $\kb^i$ is the null vector which generates our deformation, and $\lb^i$ is the other linearly-independent orthogonal null vector, normalized so that $\lb^i\kb_i =1$. We now define the null vectors $\bar{k}^\mu$ and $\bar{l}^\mu$ in the bulk which are orthogonal to the extremal surface and limit to $\kb^i$ and $\lb^i$, respecticely, as $z\to 0$. $\bar{k}^\mu$ and $\bar{l}^\mu$ can be expanded in $z$ just like $\bar{X}^\mu$, and the expansion coefficients for $\bar{k}^\mu$ and $\bar{l}^\mu$ can be solved for in terms of those for $\bar{X}^\mu$. We will perform this expansion explicitly in the next section.

Once we have constructed $\bar{k}^\mu$ and $\bar{l}^\mu$, we write
\be
s^\mu = \alpha \bar{k}^\mu + \beta \bar{l}^\mu.
\ee
The coefficients $\alpha$ and $\beta$ are determined by the requirement that $s^\mu$ be tangent to $\mK$. This is achieved by setting
\be
\alpha = g_{\mu\nu} \bar{l}^\mu \partial_\lambda \bar{X}^\nu, \quad \beta = g_{\mu\nu} \bar{k}^\mu \partial_\lambda \bar{X}^\nu.
\ee
Then the inequality $g_{\mu \nu} s^\mu s^\nu \geq 0$ becomes
\be\label{eq-alphabeta}
\alpha\beta \geq 0.
\ee
Now, $\partial_\lambda \bar{X}^\mu \to \delta^\mu_i \kb^i$ as $z\to 0$, which implies that $\alpha \to 1$ and $\beta \to 0$ in that limit. This means that the coefficient of the most slowly-decaying term of $\beta$ is non-negative. Below we will compute $g_{\mu\nu} \bar{k}^\mu \partial_\lambda \bar{X}^\nu$ perturbatively in $z$ to derive the QNEC.

\subsection{Derivation of the QNEC}\label{sec-der}

In this section we derive the QNEC by explicitly constructing a perturbative expansion for the null vector field $\bar{k}^\mu$ orthogonal to the extremal surface and compute $g_{\mu\nu} \bar{k}^\mu \partial_\lambda \bar{X}^\nu$. This requires knowledge of the asymptotic expansion of the embedding functions $\bar{X}^\mu(y,z)$ and the metric $g_{\mu\nu}$ up to the order $z^d$. Using the assumption $\kb_i K^i_{ab}=0$, which we imposed to eliminate divergences in the entropy, we have the simple expression \eqref{eq-Xexpansion} for $\bar{X}^i$, which we reproduce here,
\be\label{eq-Xexpansion2}
\bar{X}^i(y,z) = \Xb^i(y) + B(y,z) \kb^i(y) + \frac{1}{d}V^i (y)z^d + o(z^d),
\ee 
it is straightforward to construct the vector $\bar{k}^\mu$. We use the ansatz
\be
\bar{k}^\mu(y^a,z) = \delta^\mu_z k^z(y^a,z) + \delta^\mu_i \left(\kb^i(y^a) + z^d \Delta k^i(y^a)\right),
\ee
where
\be
(k^z)^2 + \left(\frac{16\pi G_N}{d L^{d-1}}T_{kk} + 2 \kb_i \Delta k^i\right)z^d = o(z^d)
\ee
ensures that $\bar{k}^\mu$ is null to the required order. We demand that $\bar{k}^\mu$ is orthogonal to both $\partial_a \bar{X}^\mu$ and $\partial_z \bar{X}^\mu$, which for $d>2$ results in the two conditions
\begin{align}
0 &= \partial_a \Xb_i \Delta k^i   + \frac{1}{d} \kb_i \partial_a V^i + \frac{16\pi G_N}{d L^{d-1}}t_{ij}\partial_a \Xb^i \kb^j,  \\
0 &= \kb_i \Delta k^i + \frac{8\pi G_N}{d L^{d-1}}T_{kk}.
\end{align}
For $d=2$ we instead have
\be
0 = \kb_i \Delta k^i + \frac{1}{2}(\kb_i V^i)^2  + \frac{4\pi G_N}{L}T_{kk}.
\ee
Together these equations determine $\Delta k^i$ up to the addition of a term proportional to $\kb^i$. This freedom in $\Delta k^i$ is an expected consequence of the non-uniqueness of $\bar{k}^\mu$, but the inequality we derive is independent of this freedom. Notice that the function $B$ plays no role in defining $\bar{k}^\mu$. This is because we are only ever evaluating our expressions up to order $z^d$, and since $\kb^i$ is null and orthogonal to $\Sigma$ there are no available vectors at low enough order to contract with $B\kb^i$ which could give a nonzero contribution.

Now we take the inner product of $\bar{k}^\mu$ with $\partial_\lambda \bar{X}^\mu$ to get
\be
g_{\mu\nu} \bar{k}^\mu \partial_\lambda \bar{X}^\nu =\left(\kb_i \Delta k^i +\frac{1}{d} \kb_i \partial_\lambda V^i +  \frac{16\pi G_N}{d L^{d-1}} T_{kk} \right)z^d + o(z^d).
\ee
Here we used the geodesic equation, $\partial_\lambda \kb^i =0$, in order to find once more that the $B\kb^i$ term in \eqref{eq-Xexpansion2} drops out. Using our constraint on $\Delta k^i$ and the inequality \eqref{eq-alphabeta} gives us the inequality
\be\label{eq-norm1}
\frac{8\pi G_N}{L^{d-1}}T_{kk} \geq  - \kb_i \partial_\lambda V^i
\ee
for $d>2$ and the inequality
\be\label{eq-norm2}
\frac{8\pi G_N}{L} T_{kk} \geq -\kb_i \partial_\lambda V^i  + (\kb_i V^i)^2
\ee
for $d=2$.

The RHS of these equations can be related to variations of the entropy using \eqref{eq-areavariation}, which we reproduce here:
\begin{align*}
\frac{\delta A}{\delta \Xb^i(y)} = C(y,\epsilon) \kb_i(y) -L^{d-1} \sqrt{\hb(y)}\ V_i(y).
\end{align*}
To convert from extremal surface area to the entropy we only need to divide by $4G_N$. Then applying \eqref{eq-areavariation} to \eqref{eq-norm1} and \eqref{eq-norm2} immediately yields
\be\label{QNEC1}
T_{kk} \geq   \frac{1}{2\pi \sqrt{\hb} } \kb^i \frac{D}{D\lambda} \frac{\delta S}{\delta \Xb^i}
\ee
for $d>2$ and
\be\label{QNEC2}
T_{kk} \geq\frac{1}{2\pi}\left[ \kb^i \frac{D}{D\lambda} \frac{\delta S}{\delta \Xb^i} + \frac{4G_N}{ L} \left( \kb^i \frac{\delta S}{\delta \Xb^i}\right)^2\right]
\ee
for $d=2$. The explicit factor $4G_N/L$ should be re-interpreted in the field theory language in terms of the number of degrees of freedom. For a CFT, we have $4G_N/L = 6/c$. When a relevant deformation is turned on, we have to use the central charge associated with the ultraviolet fixed point, $c_{UV}$. This is the appropriate quantity because our derivation takes place in the asymptotic near-boundary geometry, which is dual to the UV of the theory. In other words, $L$ here refers to the effective AdS length in the near-boundary region.

To complete the proof, we can simply restrict the support of $\kb^i$ to an infinitesimal neighborhood of the point $y$, in which case we have
\be
\kb^i(y)\frac{D}{D\lambda} \frac{\delta S}{\delta \Xb^i(y)} \to  S''(y),
\ee
where we recall the definition \eqref{eq-secondvariation} of $S''$. Then \eqref{QNEC1} and \eqref{QNEC2} imply the advertised forms of the QNEC, \eqref{QNEC}:
\begin{align*}
T_{kk} \geq \frac{1}{2\pi \sqrt{\hb}} S'' 
\end{align*}
in $d>2$ and \eqref{QNEC'}:
\begin{align*}
T_{kk} \geq \frac{1}{2\pi} \left[ S'' + \frac{6}{ c} \left(S'\right)^2\right]
\end{align*}
in $d=2$ dimensions. Following the arguments given in Section~\ref{sec-statement}, we also have the integrated form of the QNEC, \eqref{QNECglobal}:
\begin{align*}
2\pi \int dy \sqrt{\hb}~ T_{kk}  \geq \int dy ~ S''  \geq  \frac{D^2S}{D\lambda^2},
\end{align*}
as well as the analogous integrated version of \eqref{QNEC'}.

\subsection{Generalizations for CFTs}\label{sec-QNECconformal}

In this section we turn off our relevant deformation, restricting to a CFT in $d>2$. Suppose we perform a Weyl transformation, sending $\eta_{ij} \to \hat{g}_{ij} = e^{2\Upsilon} \eta_{ij}$. To find a new inequality valid for the new conformal frame, we can simply take the QNEC, \eqref{QNEC},
\begin{align*}
T_{kk} \geq \frac{1}{2\pi\sqrt{\hb}}S'',
\end{align*}
and apply the Weyl transformation laws to $T_{kk}$ and $S''$.

The effect of the Weyl transformation on $T_{ij}$ is well-known. In odd dimensions, it transforms covariantly with weight $d-2$, while in even dimensions there is an anomalous additive shift for Weyl transformations that are not part of the global conformal group. In general then
\be
T_{ij} = e^{(d-2)\Upsilon}\left(\hat{T}_{ij} - \mathcal{A}_{ij}^{(T)}\right),
\ee
where $\mathcal{A}_{ij}^{(T)}$ is the anomaly which depends on \(\Upsilon\) \cite{Cappelli:1988vw}.

The effect of the Weyl transformation on the entropy is entirely encoded in the cutoff dependence of the divergent terms. This is especially clear in the holographic context: a Weyl transformation is simply a change of coordinates in the bulk, so the extremal surface $m$ is the same before and after. The only difference is that we now regulate the IR divergences by terminating the surface on $\hat{z} = \epsilon$ with a new coordinate $\hat{z}$. Graham and Witten considered the transformation of such surface variables under Weyl transformations \cite{Graham:1999pm}. The divergent parts all transform with different weights (and shifts), so the transformation of $S$ as a whole is complicated. But the QNEC already isolates the finite part of the entropy, $S_{\rm fin}$, so we need only ask how it transforms. Graham and Witten have shown that $S_{\rm fin}$ is invariant when $d$ is odd and has an anomalous shift when $d$ is even \cite{Graham:1999pm}:
\be
S_{\rm fin} = \hat{S}_{\rm fin} - \mathcal{A}^{(S)}.
\ee
The anomalous shift $ \mathcal{A}^{(S)}$ depends on the surface $\Sigma$ as well as \(\Upsilon\), and will generically be nonzero even when $\mathcal{A}_{ij}^{(T)}$ vanishes. For a surface with $\kb_i K^i_{ab}=0$ prior to the Weyl transformation,  the anomaly is \cite{Graham:1999pm}
\be
\mathcal{A}^{(S)} = \frac{1}{8}\int dy \sqrt{\hb} \left[ \hat{K}^i \hat{K}_i   + 2 \partial_a\Upsilon \partial^a\Upsilon   \right].
\ee

Finally, we must say how $S_{\rm fin}''$ transforms. These derivatives are with respect to the affine parameter $\lambda$ which labels the flow along the geodesics generated by $\kb^{i}$. The vector tangent to the same geodesic but affinely-parametrized with respect to the new metric is $\hat{\kb}^i = e^{-2\Upsilon} \kb^i$. Acting on a scalar function $S$, the second derivative operator becomes
\be
\kb^i \partial_i \left(\kb^j \partial_jS\right) = e^{2\Upsilon}\hat{\kb}^i \partial_i \left(e^{2\Upsilon} \hat{\kb}^j \partial_j S\right) = e^{4\Upsilon}\left(\hat{\kb}^i \partial_i \left(\hat{\kb}^j \partial_j S\right)  +2 (\hat{\kb}^i \partial_i \Upsilon)(  \hat{\kb}^j \partial_j S  )\right)
\ee
Then we have, in total,
\be
S_{\rm fin}'' = e^{4\Upsilon}\left[(\hat{S}_{\rm fin} - \mathcal{A}^{(S)} )'' +2 (\hat{\kb}^i \partial_i \Upsilon)(  \hat{S}_{\rm fin}  - \mathcal{A}^{(S)})' \right],
\ee
where on the right-hand side we are careful to compute derivatives using the correctly-normalized $\hat{\kb}^i$. We also note that the expansion in the $\hat{\kb}^{i}$ direction is no longer zero after Weyl transformation, and is instead given by
\be
\hat{\theta} = \hat{\kb}^i \partial_i \log \sqrt{\hat{\hb}} = (d-2) \hat{\kb}^{i} \partial_i \Upsilon.
\ee
Putting these equations together, and dropping hats on the variables, we find that for metrics of the form $e^{2\Upsilon}\eta_{ij}$ we have a ``conformal QNEC'':
\be\label{QNECWeyl}
T_{kk} - \mathcal{A}_{kk}^{(T)} = \frac{1}{2\pi\sqrt{\hb}}\left[(S_{\rm fin} - \mathcal{A}^{(S)} )'' +\frac{2}{d-2}\theta(S_{\rm fin}  - \mathcal{A}^{(S)})' \right].
\ee
This is a local inequality that applies to all surfaces $\Sigma$ which are shearless in the \(\kb^{i}\) direction. This bound can of course be integrated to yield an inequality corresponding to finite deformations.

\subsubsection{Special case: spherical entangling regions}

The entanglement entropy across spheres has special properties compared to regions with less symmetry. Spheres minimize the entanglement entropy among all continuously-connected shapes with the same entangling surface area \cite{Astaneh:2014uba,Allais:2014ata}, which has led to the entropy of a sphere being used as a c-function \cite{Casini:2012ei,Myers:2010tj,Casini:2004bw,Komargodski:2011vj}. Spheres also play a special role because the form of their modular Hamiltonian is known explicitly \cite{Casini:2011kv,Jacobson:2015hqa,Faulkner:2013ica}

Spheres are special in the context of our analysis as well. Consider the integrated version of the conformal QNEC \eqref{QNECWeyl} specialized to the case where $\Sigma$ is a sphere in flat space. This can be obtained by a special conformal transformation from a planar entangling region (so \(\mathcal{A}_{kk}^{(T)} = 0\)). We will also choose $\kb^i$ to be uniform and directed radially inward around the sphere, so that $\theta = - (d-2)/R$, where $R$ is the sphere radius. Then we have the inequality
\be \label{sphereIntegrated}
2\pi R^{d-2} \int d\Omega \,T_{kk}(\Omega) \geq \frac{D^2}{D\lambda^2}(S_{\rm fin} - \mathcal{A}^{(S)} ) -  \frac{2}{R}\frac{D}{D\lambda}(S_{\rm fin}  - \mathcal{A}^{(S)})
\ee
For this setup, we also know that the QNEC should be exactly saturated in the vacuum state. This is because the extremal surface corresponding to a sphere on the boundary in vacuum AdS is just the boundary of the causal wedge, and uniformly transporting the sphere inward in a null direction just transports the extremal surface along the causal wedge. In other words, we know that $s^\mu$ is null, implying saturation of the inequality \eqref{sphereIntegrated}:\footnote{If the QNEC is saturated for a particular entangling surface, the conformal QNEC will be saturated for the conformally transformed surface. We can always think of this transformation as a passive Weyl transformation, which doesn't change the bulk geometry; \(s^{\mu}s_{\mu}\) is the same in all boundary conformal frames. So saturation of the conformal QNEC for a sphere in the vacuum is equivalent to saturation of the QNEC for a plane in the vacuum.}
\be
0 = \frac{D^2}{D\lambda^2}(S_{\rm fin,vac} - \mathcal{A}^{(S)} ) -  \frac{2}{R}\frac{D}{D\lambda}(S_{\rm fin,vac}  - \mathcal{A}^{(S)}),
\ee
where we used \(T_{kk} = 0\) in the vacuum. We could use this to compute $\mathcal{A}^{(S)}$ given the known result for $S_{\rm fin,vac}$. But we could just as easily subtract this equation from the previous inequality to obtain
\be
2\pi R^{d-2} \int d\Omega \,T_{kk}(\Omega) \geq \frac{D^2}{D\lambda^2}(S - S_{\rm vac} ) -  \frac{2}{R}\frac{D}{D\lambda}(S  - S_{vac}),
\ee
which is an inequality involving the vacuum-subtracted entropy of a sphere in an excited state of a CFT. Note that we no longer have to specify the finite piece of $S$ because the vacuum subtraction automatically cancels the divergent pieces.

\section{Discussion}\label{sec-future}

\subsection{Potential Extensions}

The structure of our proof was very simple, and we expect that a similar proof could extend the results beyond the regime of validity presented here. Let us review the key ingredients:
\begin{itemize}
\item It was important that the entropy was computable in terms of a surface observable which was an extremal value, in this case the area. This allowed us to focus on the near-boundary behavior of the surfaces as we made deformations of $\Sigma$, which is the only way we were able to have analytic control of the problem.

\item We had to know that the extremal surfaces moved in a spacelike way in the bulk as $\Sigma$ was deformed. In our specific case, theorem 17 of~\cite{Wall:2012uf} provided the rigorous proof of this fact, but as discussed in Section~\ref{sec:spacelike} this is should be a general property of the bulk entanglement wedge that is enforced by causality. Thus we expect that an analogous theorem can be proved in other contexts.

\item When we performed our near-boundary expansions of $s^\mu$ and $S$, we needed to find the appropriate cancellations down to order $z^d$, where the energy-momentum tensor of the field theory appeared. This cancellation was enforced by a simple geometric requirement on $\Sigma$, namely $\kb_i K^i_{ab} = 0$. It may have seemed miraculous that this happened in our holographic calculation, since it seemed to rely on special properties of the asymptotic expansions of the bulk metric and embedding functions. But cancellation of this type was expected and predicted from field theory arguments alone. Namely, these lower-order terms are the ones that determine the divergent parts of the entropy, and in general the divergent parts of the entropy are local geometric functionals which are state-independent. This means that a local geometric condition on $\Sigma$ should be enough to eliminate them, and all of the ``miraculous" properties we found stemmed from that.
\end{itemize}

\paragraph{Higher-Curvature Theories}
The proof given in this paper was set in the context of boundary theories dual to Einstein gravity. From the boundary theory point of view there is nothing particularly special about these theories, and thus if the QNEC is at all universal one would expect that the current proof could be modified to include higher-curvature theories in the bulk. 

Of the three points discussed above, the first is the most troubling. It is not known in general if the field theory entropy in an arbitrary higher-derivative theory of gravity is obtained by extremization of a local functional on a surface, though it has been shown for Lovelock and four-derivative gravity theories~\cite{Dong:2013qoa}. If this is not the case in general, then the proof of the QNEC would have to change dramatically for these other theories. 

\paragraph{Next Order in $1/N$}

It will likely be much more difficult to extend the proof to include finite-\(N\) corrections. Finite-\(N\) corresponds to quantum effects in the bulk. At the next order, \(N^{0}\), the inclusion of quantum effects require the addition of the bulk entanglement entropy across the extremal area surface \(m\) to the area of \(m\) when computing the boundary entropy \cite{Faulkner:2013ana}. It has been suggested that the correct procedure to all orders is to extremize the bulk generalized entropy ($A + S_{\rm bulk}$) instead of the area~\cite{Engelhardt:2014gca}, but for the first correction we can continue to determine $m$ by extremizing the area alone.

The difficulty in extending our proof to the next order is that, while the surface $m$ is still determined by extremizing a local functional, the entropy itself is not given by the value of that functional. So while we still have \eqref{eq-norm1}, which is an inequality involving $V^i$, the coefficient of the \(z^{d}\) term in the expansion of the embedding functions, we cannot identify $V^i$ with the variation of the entropy. Instead, the variation of the entropy is given by
\be
\label{FLMDeriv}
	 \kb^{i}\frac{\delta S}{\delta \Xb^{i}}  = \frac{1}{4 G_{N}}  \kb^{i}\frac{\delta A}{\delta \Xb^{i}}  +  \kb^{i}\frac{\delta S_{\text{bulk}}}{\delta \Xb^{i}} = \frac{\sqrt{\hb}}{4 G_{N}} \kb^i V_i  +  \kb^{i}\frac{\delta S_{\text{bulk}}}{\delta \Xb^{i}}.
\ee
Applying this result to \eqref{eq-norm1}, we find that a sufficient (but not necessary) condition for the QNEC to hold at order \(N^{0}\) is
\begin{align}
	\frac{D}{D\lambda} \frac{\delta{S_{\text{bulk}}}}{\delta \Xb^{i}} \kb^{i} \leq 0.
\end{align}
Intriguingly, this is almost the QNEC \emph{applied in the bulk}, except for two things. Notice that the variation $\delta S_{\text{bulk}}/\delta \Xb^{i}$ is a global variation of $S_{\rm bulk}$, not a local one. We could re-expand it in terms of a local variation integrated over all of $m$. But the variation of $m$ is spacelike over most of the surface, even though it becomes null at infinity. The integrated QNEC does not apply when the variation is spacelike in some places. We would also expect that the bulk stress tensor should play some role in any bulk entropy inequality.

\paragraph{Curved Backgrounds}
A straightforward generalization of this proof is the extension to field theories on a curved background. The main problem is that the state-independent terms in the asymptotic metric expansion would not be proportional to the metric and thus would not vanish when contracted with the deforming null vector \(\kb^{i}\). For example, for arbitrary bulk gravity theories dual to \(d=4\) CFTs the first two terms in the metric expansion read \cite{Imbimbo:1999bj}
\begin{align} \label{g2}
	g_{ij}(x,z) = g_{(0)ij} + \frac{z^{2}}{2} \left[ R_{ij} - \frac{1}{2(d-1)} R\ g_{(0)ij} \right] + \cdots,
\end{align}¥
where \(g_{(0)ij}\) is the boundary metric. The \(R_{ij}\) term will interfere with the proof if \(R_{kk} \neq 0\). But there is another aspect of the curved-background setup which may help: the geometrical condition we have to impose on $\Sigma$ to eliminate divergences is not just $\kb_i K^i = 0$. The second variation of the area law term in the entropy, for instance, is proportional to the derivative of the geometric expansion of a null geodesic congruence, $\dot{\theta}$, and by Raychaudhuri's equation this depends on $R_{kk}$. So it may be that the condition which guarantees the absence of divergences in the QNEC in a curved-background is also strong enough to deal with all the background geometric terms which can show up to ruin the proof.
\footnote{Update in version 2: Using a generalization of the method used in this paper, one can show that the QNEC holds when applied to Killing horizons of boundary theories living on arbitrary curved geometries \cite{KoellerFutureWork}.}

\paragraph{Quantum Focussing Conjecture}

We have discussed at length the restriction to surfaces satisfying $\kb_i K^i_{ab} =0$ as a way to eliminate divergences in the variation of the von Neumann entropy. But the original motivation for the QNEC, the Quantum Focussing Conjecture (QFC), was made in the context of quantum gravity, where the von Neumann entropy is finite (and is usually referred to as the generalized entropy). Instead of an area law divergence, the generalized entropy contains a term $A/4G_N$, and instead of subleading divergences there are terms involving (properly renormalized) higher curvature couplings. The QFC is an analogue of the QNEC for the generalized entropy, and simply states $S_{\rm gen}'' \leq 0$. When applied to a surface satisfying $\kb_i K_{ab}^i =0$ it reduces to the QNEC, but when applied to a surface where $\kb_i K_{ab}^i \neq 0$, it has additional terms involving the gravitational coupling constants of the theory.

Using out present method of proof, we could potentially study these additional gravitational terms, and hence prove some version of the QFC. The idea is to consider an induced gravity setup in AdS/CFT, where the field theory lives not on the asymptotic boundary but on a brane located at some finite position. As is well-known, the CFT becomes coupled to a $d$-dimensional graviton in this setup~\cite{Randall:1999vf,Randall:1999ee}. Furthermore, it has been shown that the area of an extremal surface anchored to the brane and extending into the bulk computes $S_{\rm gen}$ for the CFT+gravity theory on the brane~\cite{Bianchi:2012ev, Myers:2013lva}.

For a brane which is close to the boundary, we can essentially apply all of the methodology of our current proof to this situation. The only difference is that, since we are not taking $z\to 0$, we do not have to worry about setting $\kb_i K^i_{ab} = 0$ to kill the divergences. And when we compute $s_\mu s^\mu$ without the condition $\kb_i K_{ab}^i = 0$, there will be additional terms that would have dominated in the $z\to 0$ limit. Schematically, we will have
\be
0 \leq s^\mu s_\mu = z^2\left(\text{non-vanishing when $\kb_i K_{ab}^i \neq 0$}\right) + \cdots + z^d\left(T_{kk} - S''\right).
\ee
Since $z$ is left finite and is related to the finite gravitational constant of the braneworld gravity, these terms have exactly the expected form of terms in the QFC. It remains to be seen if the QFC as conjectured is correct, or if there are other corrections to it. This method should tell us the answer either way, and we will investigate it in future work.

\subsection{Connections to Other Work}

\paragraph{Relation to studies of shape-dependence of entanglement entropy}
The shape-dependence of entanglement entropy in the vacuum state of a quantum field theory has recently been an active area of research.\footnote{See e.g. \cite{Rosenhaus:2014woa,Rosenhaus:2014zza,Carmi:2015dla,Faulkner:2015csl,Allais:2014ata}.} Recent studies have focused on the explicit calculation of the ``off-diagonal'' parts of the second variation of the entropy, sometimes known as the ``entanglement density'' \cite{Nozaki:2013wia, Nozaki:2013vta, Bhattacharya:2014vja}. These terms play no role in the local version of the QNEC, which only involves the diagonal part. For an integrated version of the QNEC, it is sufficient that the off-diagonal terms are negative, a result which can be proven via strong subadditivity alone, as discussed above \cite{Bousso:2015mna,Bhattacharya:2014vja}. It would be interesting to see if any of the methods applied to the study of the entanglement density could be applied to the diagonal part of the second variation to study the QNEC for interacting theories without using holography.

\paragraph{Other energy conditions}

A number of non-local conditions on the stress tensor in quantum field theory have been suggested over the years, some more exotic than others. These include the average null energy condition (ANEC) \cite{HawEll}, as well as the more recent ``quantum inequalities'' (QIs) \cite{Ford:1994bj,Ford:1996er} which imply the ``quantum interest conjecture'' \cite{Ford:1999qv}. The motivation for non-local energy conditions in quantum field theory naturally comes from the fact that quantum fields violate all local energy conditions defined at a single point \cite{Epstein:1965zza}. 

It would be interesting to understand the relation between these inequalities, and to see which ones imply or are implied by the others. It was pointed out in \cite{Ford:1994bj} that the QIs imply the ANEC in Minkowski space, and by integrating the QNEC along a null generator one can obtain the ANEC in situations where the boundary term \(S'\) vanishes at early and late times \cite{Bousso:2015wca}. But does the QNEC imply a null limit of the QI?\footnote{In \cite{Ford:1994bj} it is mentioned that a QI can be derived for null geodesics for 1+1-dimensional Minkowski space, but that it is not known if an analogous statement holds in higher dimensions.} Or can the QI be shown to imply the QNEC? One might expect that the QNEC should be the more general statement, simply because of the huge freedom in the choice of region used to define the entropy.

\paragraph{Semiclassical generalizations of classical proofs from NEC \(\to\) QNEC}
Many proofs of theorems in classical gravity rely on the assumption of the Null Energy Condition (NEC) \cite{Hawking:1971tu,Bousso:2015mqa,Penrose:1964wq,HawEll,Wald,Morris:1988tu,Friedman:1993ty,Farhi:1986ty,Tipler:1976bi,Hawking:1991nk,Olum:1998mu,Visser:1998ua,Penrose:1993ud,Gao:2000ga}. In the context of AdS/CFT, the large-\(N\) limit of the boundary theory is dual to classical gravity in the bulk, and thus the NEC can be used to derive theorems about the AdS/CFT correspondence in this regime (e.g. \cite{Wall:2012uf,Gao:2000ga,Myers:2010tj,Myers:2010xs,Headrick:2014cta,Bunting:2015sfa}, as well as many others). One wonders about the fate of these results away from the strictly classical limit, because the NEC is known to be violated by quantum fields \cite{Epstein:1965zza}. 

As shown in this paper and \cite{Bousso:2015wca}, the QNEC is a generalization of the NEC which holds in several nontrivial examples of fully quantum theories. It would be interesting to try to replace the assumption of the NEC with the assumption of the QNEC to generalize classical proofs in gravity to the semi-classical regime. While the introduction of entropy into gravitational theorems may be a non-trivial modification, a similar program of replacing the NEC with the GSL for causal horizons \cite{Bekenstein:1973ur,Wall:2011hj} has already had success in various cases \cite{C:2013uza,Engelhardt:2014gca}. Replacing the NEC with the QNEC could potentially be even more powerful, as the QNEC holds at any point in spacetime without the need for a causal horizon.

\section*{Acknowledgements}
We would like to thank C.~Akers, R.~Bousso, X.~Dong, Z.~Fisher, M.~Mezei, M.~Moosa, and A.~Wall for discussions. We would also like to thank N.~Curington for help with the figures.
The work of JK and SL is supported in part by the Berkeley Center for Theoretical Physics, by the National Science Foundation (award numbers 1214644, 1316783, and 1521446), by fqxi grant RFP3-1323, and by the US Department of Energy under Contract DE-AC02-05CH11231.

\appendix
\section{Details of the Asymptotic Expansions}\label{stressTensorApp}

In this appendix we will provide a few more details about the asymptotic expansions appearing in section~\ref{sec:Expn}. Consider an Einstein-scalar field system where the scalar field $\Phi$ has mass $m^2 = \Delta(\Delta -d)$, and \(\Delta\) is the dimension of the relevant boundary operator \(\mO\). It is useful to also define $\alpha = d-\Delta$. Let us assume first that $\Delta \geq d/2$, so that $\alpha < \Delta$. This is the case for the standard quantization of the scalar field. Near $z=0$, the leading part of the field is then $\Phi \sim \phi_0 z^\alpha$, where $\phi_0$ is a constant which is proportional to the coupling constant of the relevant operator. Then the Einstein equations have a solution of the form given by \eqref{metricExpn},
\be\label{eq-metricExpn2}
	ds^{2} = \frac{L^{2}}{z^{2}} \left( dz^{2} + \left[ f(z) \eta_{ij} + \frac{16 \pi G_{N}}{dL^{d-1}} z^{d} t_{ij} \right] dx^{i} dx^{j} + o(z^{d})\right),
\ee
where $f(z)$ is state-independent and has an expansion
\begin{align}\label{fExpn}
	f(z) = 1 + \sum_{m=2}^{m\alpha \leq d} f_{(m\alpha)} z^{m\alpha}.
\end{align}
Here  $f_{(m\alpha)}$ is proportional to $\phi_0^m$. The minimal value $m=2$ corresponds to the fact that, in Einstein gravity, the metric couples quadratically to $\Phi$.

It is important for the this proof that all terms in the expansion of the metric and embedding functions of lower order than \(z^{d}\) are proportional to \(\eta_{ij}\) and \(k^{i}\), respectively. For the metric, we can see this immediately from \eqref{eq-metricExpn2} for operators with \(\Delta \geq d/2\). One has to be more careful in the case where $d/2 > \Delta > (d-2)/2$. The lower bound here represents the unitarity bound. Treatment of this case requires the alternative quantization, which means that the roles of $\alpha$ and $\Delta$ are switched \cite{Klebanov:1999tb}. In particular, it means that when we solve Einstein's equations there will be terms of order less than $z^d$ which are state-dependent:
\be
ds^2 =\frac{L^{2}}{z^{2}} \left( dz^{2} + \left[ \left(\eta_{ij} + \sum_{m=2,n=0}^{2n+m\Delta \leq d} g_{ij}^{(2n+m\Delta)} z^{2n + m\Delta}\right)  + \frac{16 \pi G_{N}}{dL^{d-1}} z^{d} t_{ij} \right] dx^{i} dx^{j} + o(z^{d})\right).
\ee
Here the $g_{ij}^{(2n+m\alpha)}$ are built out of the expectation value of the relevant operator, $\langle \mathcal{O} \rangle$, rather than its coupling constant. However, all is not lost. Because $\Delta > (d-2)/2$, only the coefficients with $n =0 $ actually appear in this sum because the others are $o(z^d)$.\footnote{We are excluding operators which saturate the unitarity bound, $\Delta = (d-2)/2$, because those are expected to be free scalars not coupled to the rest of the system.} But $g_{ij}^{(m\Delta)}$ depends only on $\langle \mathcal{O} \rangle^m$ and not any of its derivatives (this follows from a scaling argument~\cite{Hung:2011ta}). So $g_{ij}^{(m\Delta)} \propto \eta_{ij}$, which is what we need for the argument in the main text.

We also have to make sure that derivatives of $\langle \mathcal{O} \rangle$ do not contaminate the expansion of the embedding functions $\bar{X}^i$. From the equation of motion, we see that the lowest order at which $\partial_a \langle \mathcal{O} \rangle$ enters the expansion of $\bar{X}^i$ is $z^{2 + 2\Delta}$, but $2 + 2\Delta > d$ for $\Delta > (d-2)/2$.

%

\bibliographystyle{utcaps}
\bibliography{all}

\end{document}